# A Study of Position Bias in Digital Library Recommender Systems

Andrew Collins[1,2], Dominika Tkaczyk[1], Akiko Aizawa[2], and Joeran Beel[1,2]

[1] Trinity College Dublin, School of Computer Science and Statistics, ADAPT Centre, Ireland
{Joeran.Beel|Andrew.Collins|Dominika.Tkaczyk}@adaptcentre.ie

[2] National Institute of Informatics (NII), Tokyo, Japan
aizawa@nii.ac.jp

**Abstract.** "Position bias" describes the tendency of users to interact with items on top of a list with higher probability than with items at a lower position in the list, regardless of the items' actual relevance. In the domain of recommender systems, particularly recommender systems in digital libraries, position bias has received little attention. We conduct a study in a real-world recommender system that delivered ten million related-article recommendations to the users of the digital library Sowiport, and the reference manager JabRef. Recommendations were randomly chosen to be shuffled or non-shuffled, and we compared click-through rate (CTR) for each rank of the recommendations. According to our analysis, the CTR for the highest rank in the case of Sowiport is 53% higher than expected in a hypothetical non-biased situation (0.189% vs. 0.123%). Similarly, in the case of Jabref the highest rank received a CTR of 1.276%, which is 87% higher than expected (0.683%). A chi-squared test confirms the strong relationship between the rank of the recommendation shown to the user and whether the user decided to click it ($p < 0.01$ for both Jabref and Sowiport). Our study confirms the findings from other domains, that recommendations in the top positions are more often clicked, regardless of their actual relevance.

## 1 Introduction

Position bias is a commonly observed phenomenon in Information Retrieval. It describes a tendency of people to notice or interact with items in certain positions of lists with higher probability, regardless of the items' actual relevance. Eye tracking studies demonstrate that users are less likely to look at lower ranking items in vertical lists, typically only examining the first few entries [11]. Furthermore, 65% of users interact

---

This publication has emanated from research conducted with the financial support of Science Foundation Ireland (SFI) under Grant Number 13/RC/2106. This work was also supported by a fellowship within the postdoc-program of the German Academic Exchange Service (DAAD).

with lists in a depth-first fashion, clicking on the first item which seems relevant, without evaluating the entire list in a holistic fashion [14].

Position bias creates challenges in evaluating recommender systems based on users' interaction with recommendations. The relevance of sets of recommendations is often implicitly inferred by tracking clicks by a user on a set. The effectiveness of a recommender system can therefore be evaluated using this click data. Due to position bias, however, the probability of a user interacting with an item might not indicate that item's absolute relevance within the set. Evaluations which rely on click data, but which don't take bias into account, may be misleading.

There is little research that tests for the existence of position bias in recommender systems, and no research on position bias in recommender systems for digital libraries, to the best of our knowledge. Recommender system studies do not usually assess bias using click data from typical real-world system, and particularly not with click data which reflects typical digital library usage. A small number of user studies and offline evaluations exist, however it is not certain that the results from offline studies will be generally applicable to real-world digital library recommender systems [4]. Consequently, it remains uncertain if, and to what extent, position bias exists for recommender systems in the real-world.

Our research goal is therefore to assess if position bias exists in real-world recommender systems, in the context of digital libraries.

## 2 Related Work

Existing research into position bias in recommender systems tends to either: test for its existence through small user studies [21][23], model or simulate biased user behavior based on past data [8][19], or account for biased click behaviour with the possible goal of training a system [18][12][17][22]. In search engine research, robust eye tracking studies have also been conducted which assess its effects [11][10].

A common approach to testing position bias without the need for eye tracking interventions, in both recommender systems and search engine research, is to alter the order of ranked recommendations or search results in some manner. Users' interactions with altered orderings are then compared to that of non-altered orderings. Some studies randomly shuffle each set of items and compare them to non-shuffled sets [21][23][20]. Other studies re-order results in a specific, non-random way. For example, results are presented to users in a reverse ordering in several studies, and clicks on reversed sets compared to those of non-reversed sets [11][13].

In the case of both random shuffling, and reversed orderings, inferences about biased behaviour can be made from comparison to the correct ordering. For example, if users click the first rank of a list with similar probability when it has either a highly relevant item, or irrelevant item, position bias may be evident. Keane et al. [13] and Joachims et al. [11] each used reversed-set comparisons and found that position bias was evident in search engine usage; the percentage of clicks on the highest rank remained higher than the lowest rank with reversed rankings (40% vs 10% [13], 15% vs 5% [11]).

However in both cases, the percentage of clicks on the lowest ranking items increased in the reversed state (0% to 10% [13], 2% to 5% [11]). This suggests that, despite the effects of position bias, some users are perhaps systematic in their browsing behaviour, and will examine lists of results more thoroughly before clicking on an item.

Results from recommender system user studies are inconsistent with findings from search engine research however. Through random shuffling, Teppan and Zanker found that the position of an item in a recommender system is less important that the desirability of an item to a user, as assessed by clicks, specifically when encouraged to examine lists closely [21]. Zheng et al. found that recommendation relevance was the sole determinant of click rates, and that position bias had *no* impact on behaviour [23].

Position bias is not only tested through shuffling recommendations, but its negative effects can also be countered by taking advantage of it. Pandey et al. promote random items to the top of recommendation sets to account for position bias, which acts against new items in systems which make recommendations based on item popularity [17].

Finally, reordering has been recently used to estimate the strength of position bias in a given scenario, and derive click propensity scores for ranks within recommendation sets. These propensity scores can then be used to train performant learning-to-rank using biased feedback [12][22].

Joachims et al. suggest that the effect of position bias may be more of a problem in systems which do not assess user judgments of relevance explicitly [11]. Implicit data such as clicks on recommendations are used to approximate relevance for a user, because they are cheap to collect and analyse. For such approximation to be useful, however, there must be strong correlation between the item's relevance and clicks. The effects of position bias, therefore, may be an important consideration for digital libraries that employ recommender systems but do not encourage explicit ratings of recommendations by users.

## 3 Methodology

In order to assess position bias in recommender systems, we examined data from the digital library Sowiport [7], and reference manager Jabref [6]. Both Sowiport and Jabref use Mr. DLib, a recommendation-as-a-service provider, to recommend documents to users [2][6][3]. Sets of recommendations were chosen and ranked by the Mr. DLib recommender system based upon users' actions on Sowiport and in Jabref. The recommendations are presented to users in a vertical list format, and subsequent user-interactions with each set are tracked [1].

In total, approximately 1.6m sets, each containing 6 recommendations, were delivered to users during the course of the study. The study was run over a period of 5 months beginning in March 2017. In total 10m recommendations were delivered to users and 12,543 clicks were logged. Click-through Rates (**CTR**) – the ratio of clicked recommendations to delivered recommendations – were established for each rank of the vertical lists of recommendations. "Highly ranked" items are those that appear towards the top of lists (Figure 1).

Of the sets of recommendations which receive clicks, most receive just one click (Figure 2). We expect that any tendency for users to pay more attention to highly ranked items in a set should manifest as a disproportionately high number of clicks for those ranks on average, when compared to lower ranks.

Of the 10 million recommendations delivered, approximately half were delivered in sets whose rankings were shuffled before being presented to users. Furthermore, 1% of recommendations were chosen randomly from the entire corpus, to act as a baseline.

Due to the large amount of randomly shuffled data available, it was possible to retrospectively conduct several analyses of the click data, based upon other author's analyses.

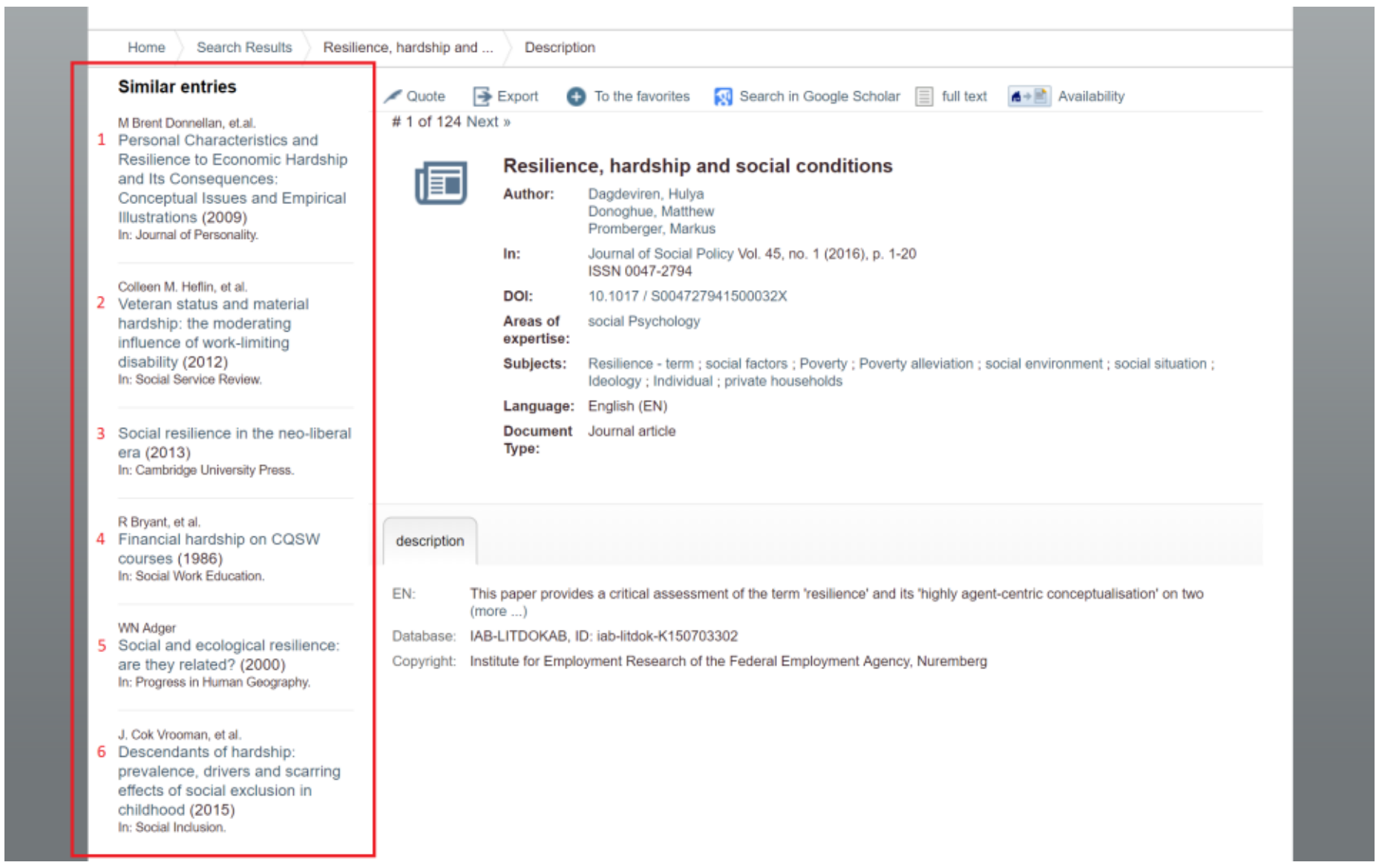

**Figure 1:** Recommendations displayed for a Sowiport document. The set of recommendations is highlighted for this figure, with each rank of the list numbered

The first analysis aims to verify the results of user studies that test for bias through random shuffling of sets of recommendations [21][23]. To do this we simply compared the total average CTR for each rank in non-shuffled data, to that of shuffled data. A Chi-squared test is used to determine if, in the shuffled data, there is a significant relationship between the rank of the recommendation and whether it was clicked by the user. A significant relationship suggests the existence of position bias. Shuffling sets of recommendations should not drastically affect click rates; all sets in this experiment were limited to six items and all items within each set will typically be somewhat relevant.

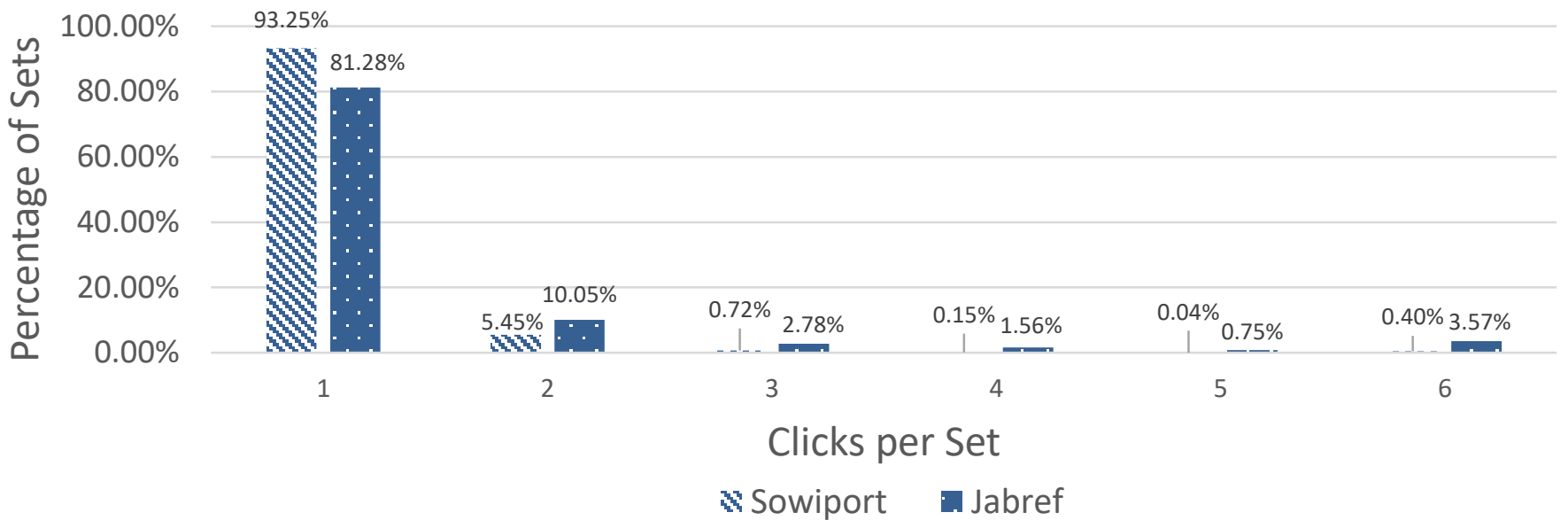

**Figure 2:** A vast majority of recommendation sets that receive at least one click, receive only one click (93% of Sowiport sets, 81% of Jabref sets)

A post hoc examination showed that ~40,000 randomly shuffled recommendations were shuffled into an approximately reversed ordering within their respective sets. Consequently, we also want to verify the results of offline studies that evaluate bias through purposeful perturbations of items presented to users, specifically, studies which compare reversed sets to non-reversed sets [11][13].

Lastly, the randomly shuffled data allows for an assessment of bias which is similar to propensity estimation as described by Joachims et al. [12], and by Wang et al. [22], in building their click propensity models. In our study, we aggregate the sets of randomly shuffled recommendations according to the rank, in which the user saw the most relevant item. We analyse the CTRs for these aggregated subsets separately.

All data of our study is be available at http://data.mr-dlib.org to enable other researchers to replicate our calculations, and use the data for extended analyses beyond the results we present.

# 4 Results

Click-through rates for non-shuffled sets appear as would be expected in an appropriate recommender system. The highest ranks experience the highest CTR (0.243% on average for Sowiport, 1.281% for Jabref), with a decreasing CTR by rank (Figure 3). Users *should* expect that an effective recommender system is delivering results in a sequentially relevant manner, so it is not surprising to see a decreasing CTR by rank when the system delivers recommendations in accordance with these expectations.

If a large number of recommendation sets are uniformly shuffled, and users assess lists in a rational, unbiased manner, it might be expected that CTR for each rank would be approximately equal. This expected CTR for each rank in shuffled sets, given unbiased behaviour on Sowiport (0.123%) and Jabref (0.683%) is shown in Figure 3 and Figure 4. It is calculated as total clicks on recommendations divided by the total number delivered. Some small user studies of position bias in recommender systems do find that, following uniform shuffling, CTR for each rank is approximately equal [23]. However in our online evaluation, the data resulting from shuffling sets of recommendations shows a significant relationship between the rank of the recommendation and whether the user decides to click it (Chi-squared test, $p < 0.01$). CTR for shuffled sets decreases at approximately the same rate as that of non-shuffled, rank to rank (Figure 3).

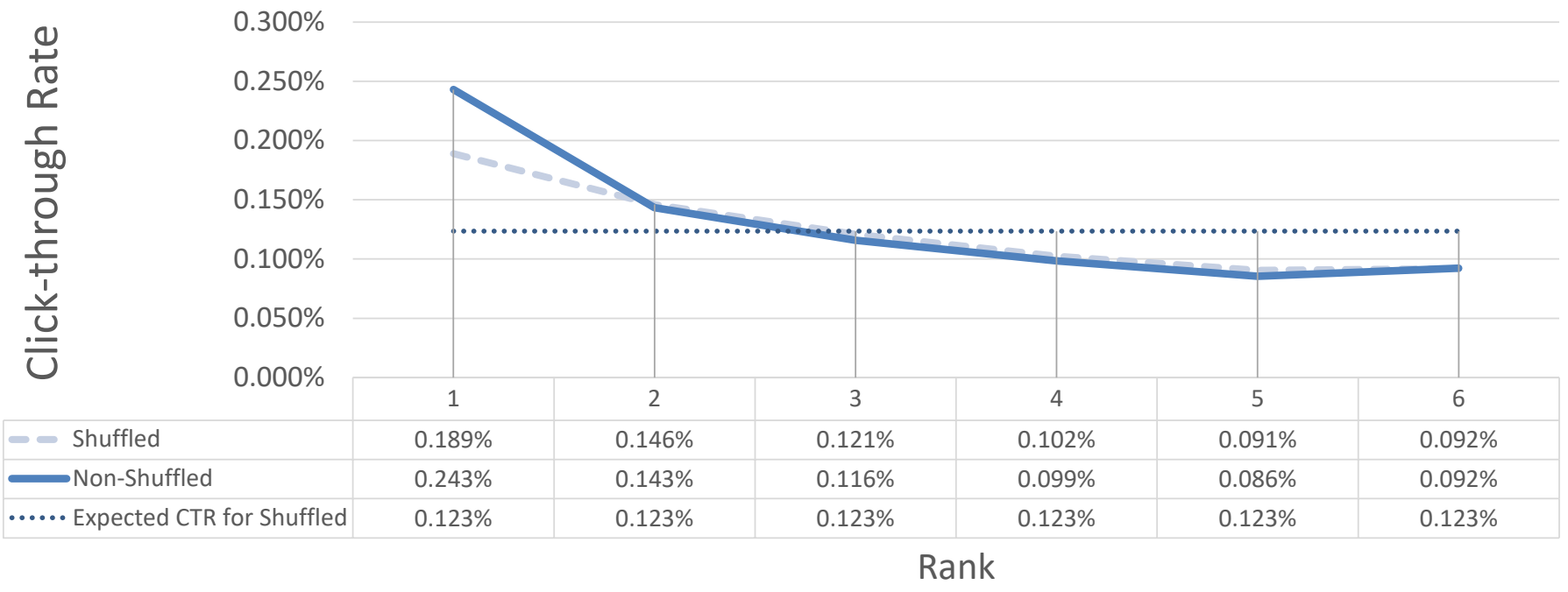

| Rank | 1 | 2 | 3 | 4 | 5 | 6 |
|---|---|---|---|---|---|---|
| Shuffled | 0.189% | 0.146% | 0.121% | 0.102% | 0.091% | 0.092% |
| Non-Shuffled | 0.243% | 0.143% | 0.116% | 0.099% | 0.086% | 0.092% |
| Expected CTR for Shuffled | 0.123% | 0.123% | 0.123% | 0.123% | 0.123% | 0.123% |

**Figure 3:** Users of the digital library Sowiport appear to exhibit bias in choosing items, following a random shuffling of recommendations

Similar results can be seen for users of Jabref (Figure 4), with a significant difference between shuffled CTR and that of a uniform distribution (Chi-squared test, $p < 0.01$).

In both cases, significant position bias seems evident in users' clicking behavior. With no meaningful difference in CTR between non-shuffled and shuffled sets on average, it seems as if users do not interact with recommendations in a rational manner.

A decreasing CTR by rank for shuffled recommendations may suggest several things. First, it may tell us that users do not care about recommendation quality and will click items regardless of their relevance simply based on their rank. Second, it may suggest that users do interact with lists in a "depth-first" manner [14], and don't assess lists holistically. That is, they may choose a poor recommendation in the second rank, because they have not noticed that there's a relatively better recommendation in e.g. rank six. Finally, it may suggest that the system is not discretely ranking recommendations well enough, and that users can see that all items are of similar relevance and, perhaps, higher ranks are more convenient to click.

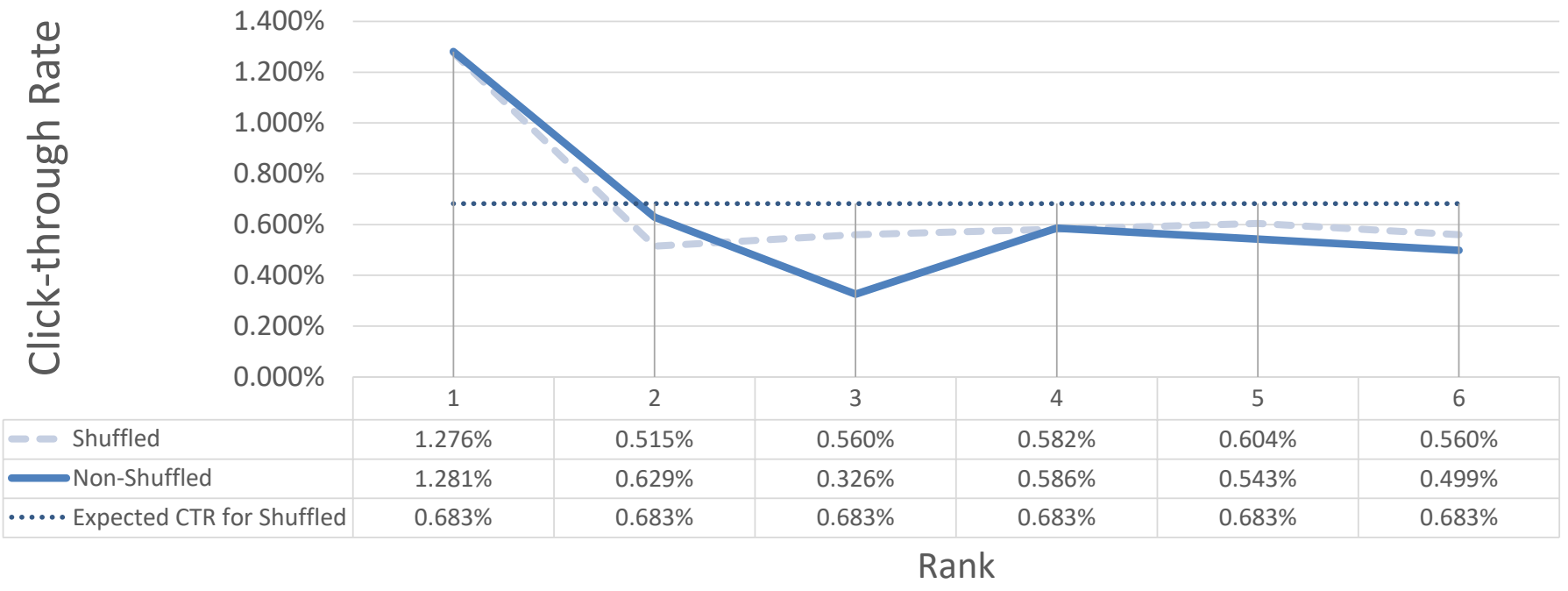

| Rank | 1 | 2 | 3 | 4 | 5 | 6 |
|---|---|---|---|---|---|---|
| Shuffled | 1.276% | 0.515% | 0.560% | 0.582% | 0.604% | 0.560% |
| Non-Shuffled | 1.281% | 0.629% | 0.326% | 0.586% | 0.543% | 0.499% |
| Expected CTR for Shuffled | 0.683% | 0.683% | 0.683% | 0.683% | 0.683% | 0.683% |

**Figure 4:** Users of reference manager Jabref also appear to exhibit position bias, following random shuffling of recommendations

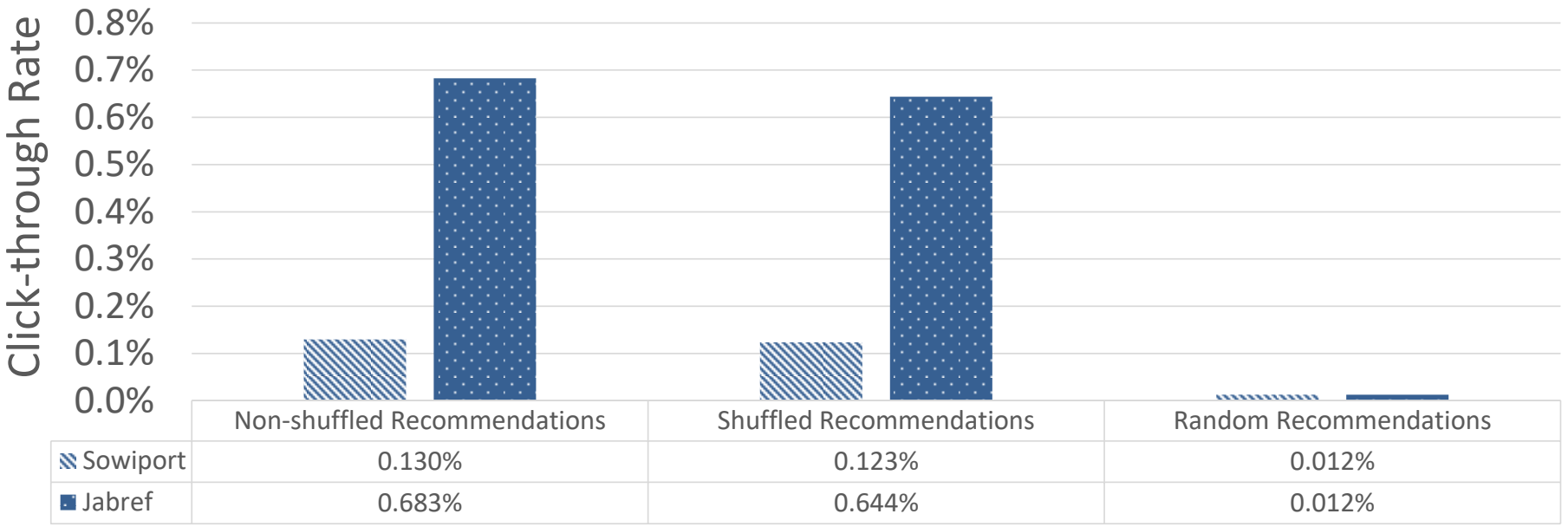

**Figure 5:** CTR for non-shuffled, shuffled, and random recommendations

The suggestion that users do not care about recommendation quality can be ruled out immediately: in the case of Mr. DLib, CTR for arbitrarily random recommendations is miniscule when compared to ranking algorithms (0.012% vs 0.130% for Sowiport, 0.012% vs 0.683% for Jabref) (Figure 5). Users are keenly aware of bad recommendations and will generally refuse to interact with them.

The second and third suggestions – “depth first” searching, and insufficiently discrete ranking – require further analysis:

When rankings are displayed in reverse ordering, the least relevant items are placed into the highest ranks. Comparing non-shuffled sets to reverse-ordered sets produces results in-line with Keane et al. [13], and Joachims et al. [11], who evaluated bias in a similar manner (Figure 6). On average, highly ranked items maintain a high CTR despite their lower relevance. Similarly, the most relevant items being placed into the lowest ranks incurs a lower CTR than when they’re left in the highest rank. As seen in the above studies, despite position bias still being evident here, the distribution of CTR seems to shift to lower ranks, with a significantly reduced CTR for the highest rank, and a significantly increased one for the lowest.

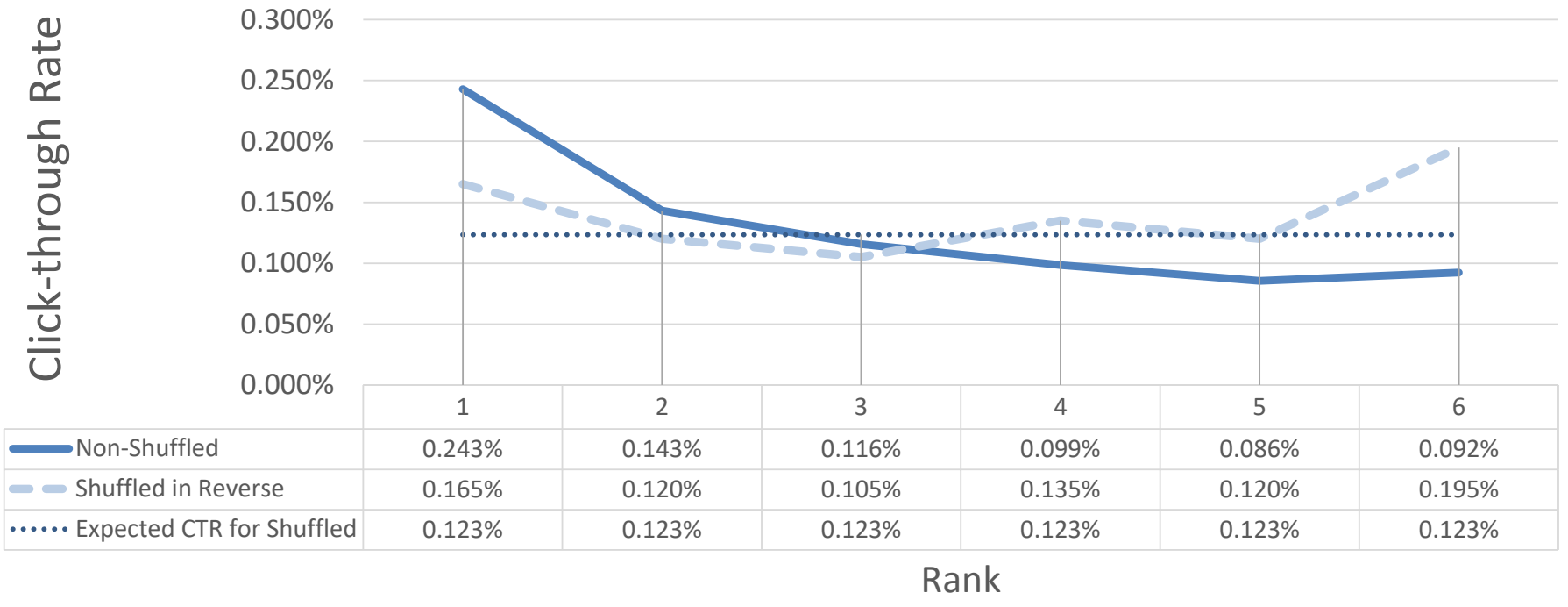

**Figure 6:** CTR by rank for non-shuffled sets on Sowiport versus sets which were shuffled into an approximately reverse ranking

This suggests that, although people are biased on average, sometimes recommendation relevance is deduced by users in a way which seems to agree with the ranking algorithm used. It is not clear whether this is due to characteristics of some subset of recommended items, which encourage searching, or characteristics of a subset of users. It is known in position bias research, for example, that users are better able to remember the first and last items in lists when compared to middle items [16]. It is also commonly seen that CTR increases for last items in lists, even when items are ordered by relevance. In Psychology research this has been explained as 'contrarian' behavior, exhibited by a small proportion of people who interact with lists from bottom to top [15]. It is unclear then whether the increased CTR for the lowest rank in reversed order is due to people seeking out relevant items throughout the entire list, or is due to contrarian interactions combined with a highly attractive item – luckily placed – in the last position. In other words, does Figure 6 show that position bias exists in recommender systems but can be somewhat overcome by excellent recommendations, or does it simply show that some subset of people exhibit position bias in reverse, and will click a relevant item if it's in the last position?

When uniformly shuffled data is analysed according to where the ranking algorithm's *most relevant* recommendation is placed, user behavior is more clear (Figure 7, Figure 8). Every analyses still seems to show position bias: higher ranks still receive a higher average CTR, and lower ranks a smaller one, despite recommendations being shuffled randomly. However, on average, regardless of rank, people seek out and are able to discern the most relevant items. CTR increases for all ranks which have the most relevant item shuffled to it, with an average increase of 29%. This suggests that increases in CTR for the lowest rank of reversed sets, as shown in Figure 6 and by Keane et al., are likely *not* due to 'contrarian' click-behavior. This seems to mirror Joachims et al.'s eye-tracking study which shows that users are more likely to examine lower ranks when presented with less relevant items in higher ones, and that they do so in a sequential fashion [11].

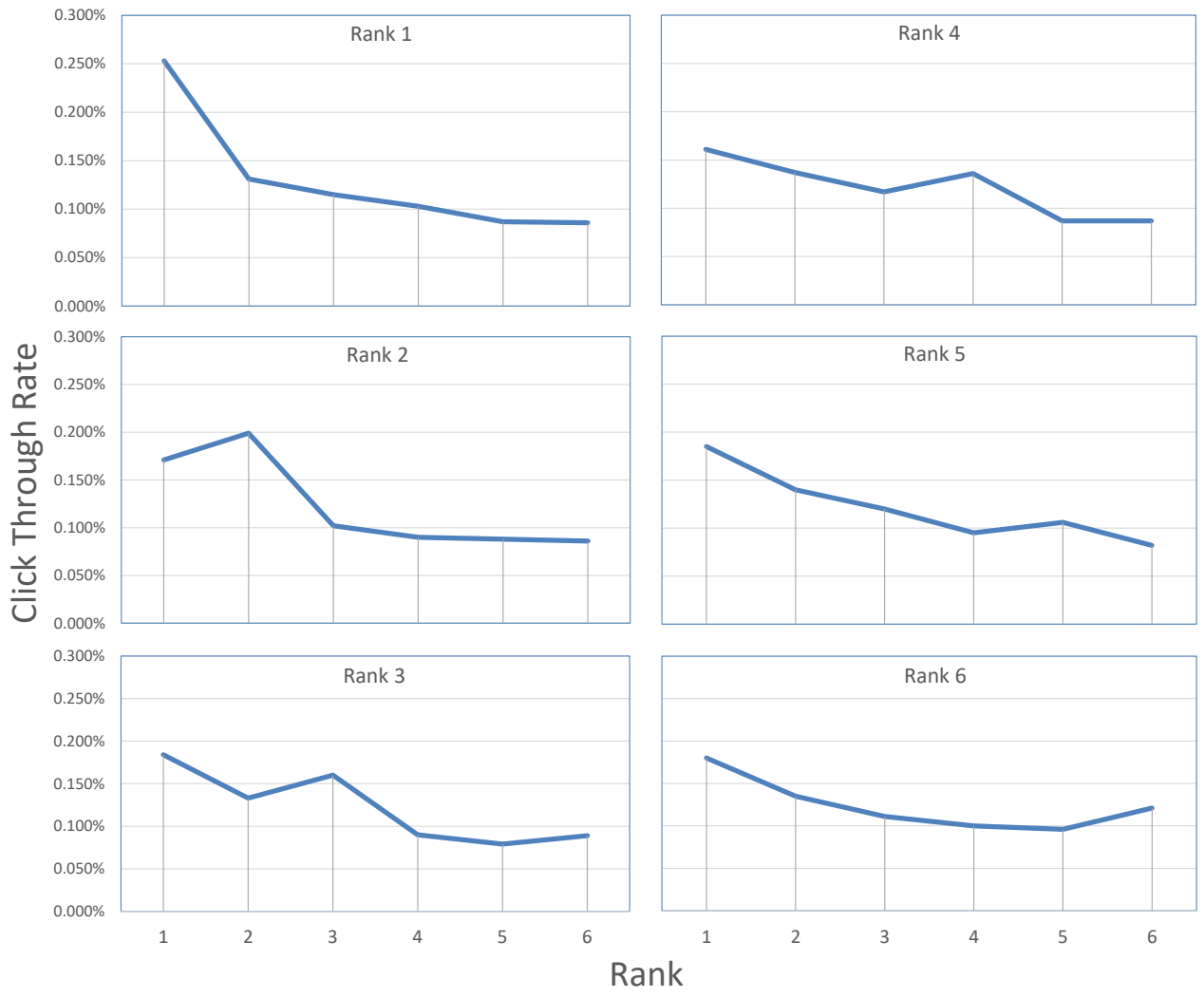

**Figure 7:** Effect on Click-Through Rate seen when aggregating shuffled Sowiport data according to where the highest recommendation has been shuffled to. For instance, the top-right graph

shows average CTR by rank when the most relevant item was randomly shuffled to rank 4 (all other ranks are randomly shuffled, too). The most relevant recommendation results in an outlying average CTR for its rank. Some portion of users seem to seek out relevant information to an equal degree at all ranks.

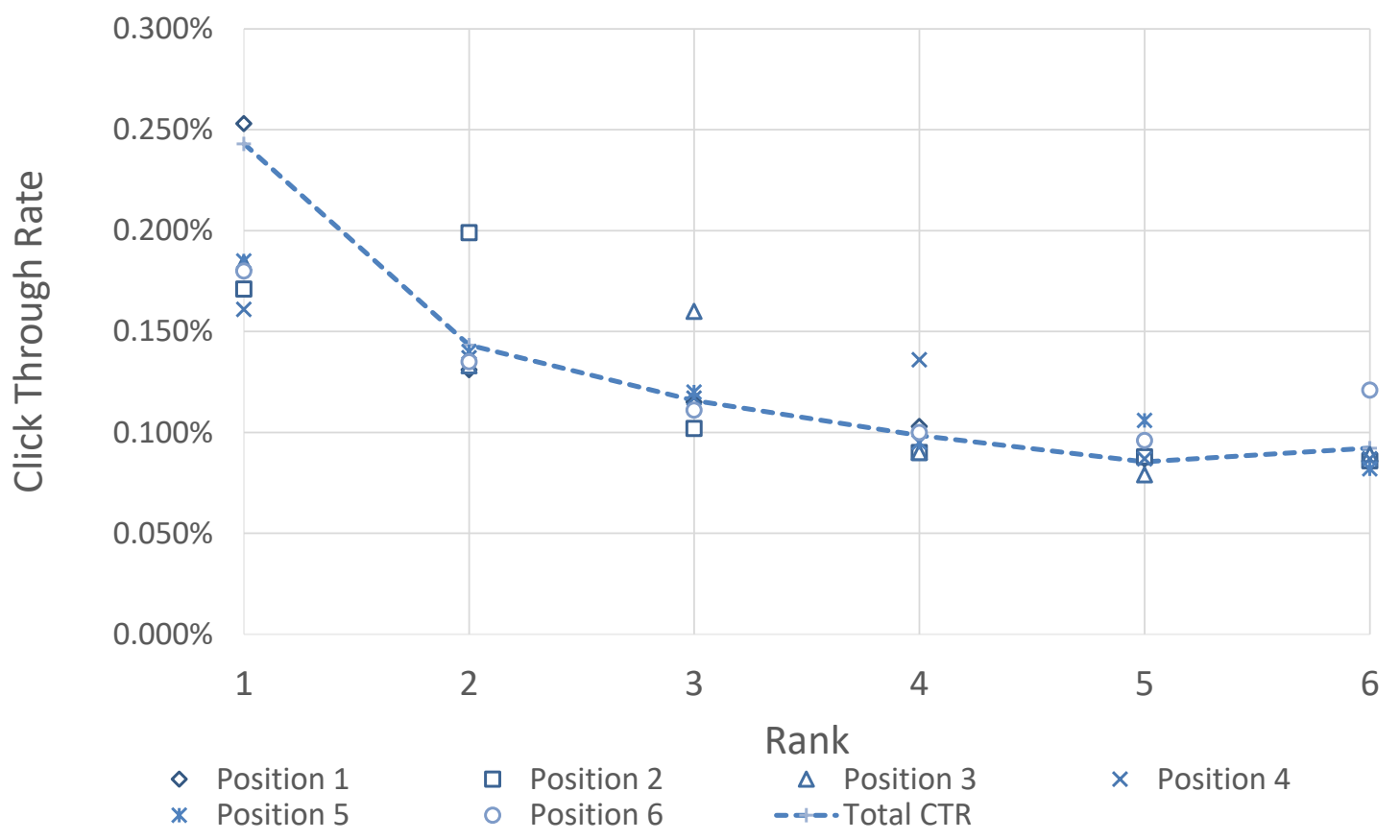

**Figure 8:** *Figure 7* combined into one graph. Each symbol represents the average CTR-by-rank for each aggregation of Sowiport shuffled data, e.g. the circle symbol shows the average CTR-by-rank for sets whose most relevant recommendation was shuffled to rank six. CTR is increased for all ranks containing the most relevant recommendation, by an average of 29.07%

# 5 Conclusion and Future Work

Our research confirms that position bias exists for recommender systems in digital libraries. The analysis shows that articles recommended at higher positions received significantly higher click-through rates than expected, regardless of their actual relevance. The CTR for the highest rank in the case of Sowiport is 0.189%, which is 53% higher than expected (0.123%). In the case of Jabref the highest rank received a CTR of 1.276%, which is 87% higher than expected (0.683%). A chi-squared test confirms the strong relationship between the rank of the recommendation shown to the user and whether the user decided to click it ($p < 0.01$ for both Jabref and Sowiport). However, our research also shows that a significant number of users look at all recommended items. Ranking recommendations in a reverse order shows click-through rates which mirror this reversal, although bias is still apparent. More distinctly, in shuffled recommendation lists, ranks containing the most relevant recommendation have a higher average CTR (29.07%) than the average CTR for shuffled recommendations on that position.

In future work, position bias could be tested in the presence of ranking and interaction biases, such popularity bias [9], and choice overload [5]. It should also be deter-mined how to quantify position bias given different modes of user interaction with recommender systems within digital libraries, as opposed to other domains. The effectiveness of unbiased learning-to-rank may then be tested for digital libraries with comparison to classical recommendation algorithms.